# Nanoparticles for Multimodal Antivascular Therapeutics: Dual Drug Release, Photothermal and Photodynamic Therapy.

Juan L. Paris[1,*], Gonzalo Villaverde[1], Sergio Gómez-Graña[1] and María Vallet-Regí[1,*].

[1] Dpto. Química en Ciencias Farmacéuticas (Unidad Docente de Química Inorgánica y Bioinorgánica), Facultad de Farmacia, Universidad Complutense de Madrid, Instituto de Investigación Sanitaria Hospital 12 de Octubre (imas12), 28040-Madrid, Spain. Centro de Investigación Biomédica en Red de Bioingeniería, Biomateriales y Nanomedicina (CIBER-BBN), Spain.

* Corresponding authors: juanluisparis@ucm.es, vallet@ucm.es

## Abstract

The poor delivery of nanoparticles to target cancer cells hinders their success in the clinical setting. In this work, an alternative target readily available for circulating nanoparticles has been selected to eliminate the need for nanoparticle penetration in the tissue: the tumor blood vessels. A tumor endothelium-targeted nanoparticle (employing an RGD-containing peptide) capable of co-delivering two anti-vascular drugs (one anti-angiogenic drug and one vascular disruption agent) is here presented. Furthermore, the nanodevice presents two additional anti-vascular capabilities upon activation by Near-Infrared light: provoking local hyperthermia (by gold nanorods in the system) and generating toxic reactive oxygen species (by the presence of a photosensitizer). RGD-targeting is shown to increase uptake by HUVEC cells, and while the nanoparticles are shown not to be toxic for these cells, upon Near-Infrared irradiation their almost complete killing is achieved. The combination of all four therapeutic modalities is then evaluated in an *ex ovo* fibrosarcoma xenograft model, which shows a significant reduction in the number of blood vessels irrigating the xenografts when the nanoparticles are present, as well as the destruction of the existing blood vessels upon irradiation. These results suggest that the combination of different anti-vascular therapeutic strategies in a single nanocarrier appears promising and should be further explored in the future.

1. Introduction

Cancer remains one of the leading causes of mortality, and recent studies estimate that it will have provoked more than 9.5 million deaths worldwide in 2018[1]. Despite many efforts to develop anti-cancer therapeutic strategies, the amount of new drugs that reach the market is very limited, and the results obtained with them is often disappointing [2]. Traditional chemotherapy is aimed towards directly killing cancer cells employing highly toxic drugs, and is currently the only available tool for many types of cancer[3]. The development of nano-drug delivery systems (nano-DDS) that could selectively deliver anticancer drugs to the tumor site was expected to transform cancer treatment, by allowing a reduction in the necessary dose of the drugs and reducing the toxic side effects accordingly[4].

Several nanomedicines have reached the market, and some areas appear as especially promising for this type of technology[5,6]. One of these areas would be combination therapy though nano-DDS that can simultaneously deliver several drugs in an optimal ratio[7–9]. This can significantly enhance the efficacy of the combined therapy, as can be seen in the recently accepted drug Vyxeos®[10]. Vyxeos® is a liposomal formulation that comprises a combination of daunorubicin and cytarabine that has been demonstrated to extend overall survival in acute myeloid leukaemia patients[10]. Despite all the achieved progress, nano-DDS for oncology have found a large amount of barriers that severely limit their efficacy[11–13]. Among these, the extremely limited penetration capacity of nanomedicines in solid tumors has been recognized as one of the most important factors hindering nanomedicine success in cancer[14–17]. A possible alternative to enable the fulfillment of nano-DDS complete potential is to direct nanomedicines towards targets that are readily available for systemically-delivered nanoparticles without the need for deep penetration into the tumor tissue. One of such targets with great potential would be endothelial cells in tumor blood vessels, developing nanomedicines for anti-vascular therapy[18,19].

In order to grow, cancer cells need the oxygen and nutrients that are provided by blood vessels. Since tumor cells without enough oxygen supply will not be able to divide, in growing tumors endothelial cells actively release many angiogenic molecules, such as EGF, TNF-α and VEGF, that induce blood vessel formation while the generation of anti-angiogenic factors is greatly decreased[20]. In anti-vascular therapy, tumors are treated by attacking the tumor blood supply, removing the oxygen and nutrient source and "starving" the cancer cells. Clinically-approved anti-angiogenic drugs (AADs, such as bevacizumab[21]) compromise the capacity of solid tumors of inducing the formation of new blood vessels, and they have shown significant efficacy against different tumor types in humans[21]. However, the effect of AADs by themselves is quite limited, and different combinations with other drugs are often used in the clinical practice. Another modality of anti-vascular therapy that is currently under evaluation is the use of vascular disrupting agents (VDAs) that can destroy already formed vessels in the tumor[22]. While VDAs appear to hold great promise as a therapeutic tool in oncology, their efficacy without combination with other agents also appears to be low, and prone to the development of resistance in the tumor.[23] The combination of AADs with VDAs has been recently proposed as a promising approach to maximize the therapeutic potential of anti-vascular therapeutics[22].

A clear synergy appears here: tumor blood vessels appear as ideal targets for nano-DDS in oncology, since there is no need for nanoparticle penetration, one of the most important limiting factors in nanomedicine for solid tumors. Moreover, anti-vascular therapy seem to rely on combinations of several drugs to be effective, a point where nanomedicine is showing an outstanding potential[18]. While nanoparticle co-delivery of an anti-angiogenic molecule plus a chemotherapeutic drug has been previously reported[24–26], to the best of our knowledge, dual delivery of two anti-vascular drugs, an AAD and a VDA, by a tumor endothelium-targeted nanodevice has not been yet described. Furthermore, due to the great versatility of nanoparticle design possibilities, additional anti-vascular modalities can be introduced in one single stimulus-responsive nanodevice, taking advantage of other therapeutic mechanisms, such as photothermal therapy (PTT, generating heat upon irradiation) and photodynamic therapy (PDT, generating reactive oxygen species, or ROS, when irradiated).

The main goal of the present work is to suggest the use of nanoparticles to perform multimodal anti-vascular therapy of solid tumors. To do this, we have developed a proof-of-concept multi-modal anti-vascular nano-DDS based on mesoporous silica nanoparticles (Figure S1). Four therapeutic modalities will be contained in this formulation: The inclusion of a gold nanorod core will provide PTT capabilities upon Near-Infrared (NIR) irradiation[27], while the coupling of the photosensitizer Indocyanine Green (ICG) will enable PDT simultaneously under the same stimulus[28]. Then, a mesoporous silica shell will allow loading and releasing our two drugs of interest[7,29–31]: an AAD (Doxycycline[32] or DOXY) and a VDA (Combretastatin A4 phosphate, also known as Fosbretabulin[33], or FOS). Finally, a polyethylene glycol (PEG) chain with an RGD-containing peptidic targeting sequence would enable the long circulation of the nanodevice in the bloodstream until reaching the target tumor endothelial cells (which are known to over-express integrins that can interact with the RGD peptidic sequence employed here[26]). A conceptual representation of the strategy proposed in this work can be seen in **Figure 1.**

## 2. Materials and Methods

### 2.1 Materials

The following compounds were purchased from Sigma-Aldrich (Spain), and were used without further purification: ammonium nitrate, tetraethyl orthosilicate (TEOS), pyridine, (3-aminopropyl)triethoxysilane (APTES), cetyltrimethylammonium bromide (CTAB), fluorescein isothiocyanate (FITC), ammonia, N,N-diisopropylethylamine (DIPEA), Indocyanine green (ICG), doxycycline (DOXY), fosbretabulin (FOS), amino-protected Fmoc aminoacids, piperidine, 2',7'-Dichlorofluorescin diacetate (H2DCFDA), N,N,N',N'-Tetramethyl-O-(1H-benzotriazol-1-yl)uronium hexafluorophosphate, O-(Benzotriazol-1-yl)-N,N,N',N'-tetramethyluronium hexafluorophosphate (HBTU), 1-Hydroxybenzotriazole hydrate (HOBt), Trifluoroacetic acid (TFA), Triisopropyl silane (TIPS), O,O'-Bis[2-(N-Succinimidyl-succinylamino)ethyl]polyethylene glycol 2kDa (NHS-PEG-NHS), Methoxypolyethylene glycol 5,000 acetic acid N-succinimidyl ester 5kDa (NHS-PEG), 2-chlorotrityl resin, Sephadex G-25, as well as the solvents used in the condensation, deprotection and release stages, such as N',N'-dimethylformamide (DMF) and dichloromethane (DCM).

Dulbecco's Modified Eagle's Medium (DMEM), Penicillin–streptomycin, and trypsin–EDTA were purchased from Invitrogen (Fisher Scientific, Spain). Endothelial growth medium (EGM) was purchased from Innoprot (Dismadel, Spain). Collagen I (Rat Tail) was purchased from Gibco. The employed cell lines (HUVEC and HT-1080 cells) were purchased from Sigma-Aldrich, and their cultures were maintained following the provided instructions.

All other chemicals (absolute ethanol, acetone, ethyl acetate, toluene, heptane, dry solvents, etc.) were of the highest commercially available quality and used as received.

### 2.2 Characterization techniques

Fourier Transformed Infrared (FTIR) spectra were obtained in a Nicolet (Thermo Fisher Scientific) Nexus spectrometer equipped with a Smart Golden Gate ATR accessory. Transmission Electron Microscopy (TEM) was carried out with a JEOL JEM 2100 instrument operated at 200 kV, equipped with a CCD camera (KeenView Camera). Scanning electron microscopy (SEM) was performed with a JEOL JSM 6335F electron microscope. X-ray diffraction (XRD) was performed using a Philips X'Pert Multipurpose Diffractometer (MPD) equipped with Cu Kα radiation. $N_2$ adsorption was carried out on a Micro- meritics ASAP 2020 instrument; surface area was obtained by applying the Brunauer–Emmett–Teller (BET) method to the isotherm and the pore size distribution was determined by the Barrett–Joyner–Halenda (BJH) method from the desorption branch of the isotherm. Mesopore diameter was obtained from the maximum of the pore size distribution curve. Z potential and hydrodynamic size of nanoparticles by Dynamic Light Scattering (DLS) were measured by means of a Zetasizer Nano ZS (Malvern Instruments) equipped with a 633 nm "red" laser. $^1$H Nuclear Magnetic Resonance (NMR) experiments were carried out in a Bruker AV 250MHz apparatus. UV–visible spectrophotometry was performed using a HELIOS-ZETA UV-vis spectrophotometer. Fluorescence spectrometry was performed by means of a Biotek Synergy 4 device. Fluorescence microscopy was performed in an Evos FL Cell Imaging System equipped with tree Led Lights Cubes ($\lambda_{Ex}$ (nm); $\lambda_{Em}$ (nm)): DAPI (357/44; 447/60), GFP (470/22; 525/50), RFP (531/40; 593/40) from AMG (Advance Microscopy Group). Quantitative analysis of cellular uptake was performed by flow cytometry in a BD FACSCalibur cytometer, and results were processed using Flowing Software. A Newport Diode Laser was used, with a continuous-wave NIR laser at 808 nm, the maximum fluence was 3 W cm$^{-2}$ and the spot size 5 mm. Pre- and postillumination temperatures of the experiments were measured by a fluorooptic probe Luxtron I652.

**2.3 Solid phase peptide synthesis**

RGDR Peptide synthesis was performed as shown in Scheme S1, and comprised the following steps: 1) *Anchoring first amino acid (Arg) to the 2-chlorotrytil resin by SN1 reaction*: Protected Arg (2 equivalents based on the mmols of the resin load) was dissolved in 3 mL of DMF, the resin was suspended in the solution and finally, DIPEA (6 eq) was added. The resulting slurry was mechanically stirred overnight. The solid was filtered and washed with DMF. 2) *Amino group Fmoc deprotection*: A solution of 20% of piperidine in DMF was added to the peptide-functionalized resin. The suspension mixture was stirred in the solid phase reactor with a wrist-shaker overnight. The solid was filtered and washed with DMF. 3) *Peptide amide bond formation*: A solution of HOBt (3 eq), HBTU (3 eq) and the amino acid (2 eq) in DMF (2 mL) was added to the washed resin after slight shaking finally DIPEA (6 eq) was added. The mixture was stirred with a wrist-shaker overnight. The solid was filtered and washed with DMF. 4) *Release of the final peptide from the resin*: Once the last amino acid was coupled and deprotected, the resin was filtered, washed with DMF and DCM. A solution of trifluoroacetic acid (95%), triisopropyl silane (2.5%) and water (1%) (2 mL) was added dropwise. The mixture was lead under mechanical stirring for 4 hours. Then, the dispersion was filtered and washed with another milliliter of the same mixture. Both filtered solutions were mixed. The crude product was obtained by precipitation of the filtered solution in cold ether. 5) *Final peptide isolation by chromatography*: The resulted solid was dissolved in the minimum amount of water and put into a flash column for molecular exclusion chromatography (stationary phase: Sephadex® G-25; mobile phase: water). All phases were frozen at -80 °C, lyophilized and characterized by NMR. The corresponding $^1$H NMR spectra can be found in Figure S2.

**2.4 Synthesis of gold nanorods** *(GNRs)*

GNRs were prepared using a modified seeded growth method.[27] Gold concentration was determined from the extinction spectra using the absorbance at 400 nm.

## 2.5 Synthesis of Mesoporous Silica-coated GNRs *(Au@MSNs)*

Coating of Au NRs with mesoporous silica (Au@MSN) was performed following a previously described protocol, with minor modifications.[27,34] Excess reactants were removed from the freshly prepared GNR solutions via two cycles of centrifugation, after which the particles were resuspended in 0.1 M CTAB, at a final gold concentration of $5 \times 10^{-3}$ M. Subsequently, 20.4 mL of a $6 \times 10^{-3}$ M CTAB solution was mixed with 60 mL of ethanol and 134 mL of water at 30 °C in a 500 mL round beaker under magnetic stirring. Upon equilibration at 30 °C for 10 min, 400 µL of $NH_4OH$ (25 vol %) was added to adjust the pH value to ≈9. Then, 6 mL of the GNR solution were added. After 5 min to ensure homogeneity of the solution, 160 µL of TEOS was added dropwise under vigorous stirring. The reaction mixture was allowed to react at 60 °C overnight. The synthesized particles were centrifuged (20 min; 9000 rpm; 35 °C), and washed in ethanol.

To obtain the final material, a series of reactions were performed, as follows (Scheme S2):

1) As-prepared Au@MSNs were functionalized with aminopropyl groups as follows: 20 mg of Au@MSNs were dispersed in 3 mL of dry toluene under $N_2$ atmosphere and placed at 80 ºC under magnetic stirring. Then, 5 µL of APTES were added and left to react at 80 ºC overnight under inert atmosphere. The particles were then collected by centrifugation and washed with toluene and ethanol (3 times). To prepare FITC-labeled nanoparticles, a similar procedure was followed, but first a reaction was performed in a different vial with 1 mg of FITC and 2.2 µL of APTES in 100 µL of ethanol with magnetic stirring at room temperature for 30 min (solution 1). Then, 10 µL of APTES were added to solution 1 and 60 µL of that mixture were added into the nanoparticle suspension in dry toluene. The rest of the procedure was performed in the same way, but protecting the FITC-labeled material from light.

2) The surfactant of amino-functionalized Au@MSNs was extracted by ionic exchange employing 10 mL of 10 mg/mL $NH_4NO_3$ solution in 95% ethanol (65 ºC, overnight, under magnetic stirring). The particles were then collected and washed with ethanol twice, obtaining Au@MSNs-$NH_2$.

3) The photosensitizer ICG was coupled to amino groups on the nanoparticle surface following a protocol described elsewhere[35] by mixing 10 mg of nanoparticles with 0.5 mL ethanol, 0.5mL pyridine and 5 µg ICG. The mixture was stirred at room temperature for 2 h. The particles were collected by centrifugation and washed three times with ethanol, obtaining Au@MSN-ICG. The ICG grafting efficiency was estimated by fluorimetry after all the washing steps.

4) Finally, the particles were PEGylated employing NHS-PEG-RGD: First, 4 mg of NHS-PEG-NHS were reacted with 1 mg of RGDR in 2 mL DMF (and with 10 µL DIPEA) under inert atmosphere at room temperature for 2 h. Then, that mixture was added to 4 mg NPs in dry DMF (total volume: 2.5 mL) and reacted overnight at room temperature under $N_2$. The particles were then collected and washed with DMF, water and ethanol, affording our desired material Au@MSN-PEG-RGD. A similar procedure was employed for non-targeted NHS-PEG maintaining the same molar ratios (preparing then Au@MSN-PEG).

## 2.6 Cargo loading

For *in vial* drug release experiments, Au@MSNs-$NH_2$ nanoparticles were initially employed, while for the final *ex ovo* evaluation of the complete system, Au@MSN-PEG-RGD particles were used. The prepared particles were loaded with our two drugs as follows: 4 mg nanoparticles were dispersed in 1 mL drug solution in PBS (10 mg/mL DOXY and 1 mg/mL FOS), stirring at room temperature overnight. The 10:1 DOXY:FOS ratio was chosen to maximize DOXY loading, since the amount of FOS introduced in the nanoparticles when co-loaded with DOXY was seen to be much larger. The particles were washed with PBS to removed non-loaded drug.

**2.7 Cargo Release**

Four mg of loaded nanoparticles were suspended in 2 mL of pH 7.4 PBS (10 mM). Then, 0.5 mL of nanoparticle suspension were placed on a Transwell® permeable support with 0.4 µm of polycarbonate membrane (3 replicas were performed in a 12 well plate). The well was filled with 1.5 mL of PBS pH 7.4 (10 mM) and the suspension was stirred at 37 ºC and 150 rpm during all the experiment. At every time point studied, the solution outside the Transwell® insert was replaced with fresh medium and the amount of each released cargo was determined by fluorescence spectrometry (DOXY: $\lambda_{exc}$ 406, $\lambda_{em}$ 515 nm; FOS: $\lambda_{exc}$ 328, $\lambda_{em}$ 400 nm).

**2.8 *In vial* evaluation of photothermal and photodynamic effect**

The capacity of the prepared nanoparticles to generate heat and ROS when exposed to NIR irradiation was first evaluated *in vial* as follows: Suspensions of Au@MSNs and Au@MSN-ICG in PBS (50 µg/mL, 1 mL total volume) previously incubated at 37 ºC were placed in wells of a 24-well culture plate and exposed to 808 nm NIR laser irradiation (1W/cm$^2$, 5 min). Temperature was measured right before and immediately after irradiation with a fluorooptic probe Luxtron I652. ROS generation was also evaluated by having the ROS indicator H2DCFDA in the medium prior to irradiation and measuring its fluorescence right after irradiation (following the manufacturer´s instructions).

**2.9 Cell culture experiments**

Cell culture experiments with Human umbilical vein endothelial cell (HUVEC) were carried out. HUVEC cells were seeded in 24 well plates at a density of 20 000 cells per cm$^2$ 24 h before the experiments were performed.

For nanoparticle uptake experiments, cells were incubated with 25 or 50 µg/mL suspension of the nanoparticles (with or without the targeting moiety) in complete endothelial growth medium (containing serum, 0.5 mL volume per well). The particles were incubated with the cells for 2 h, and then they were removed and the cells were washed with PBS to remove non-internalized nanoparticles. Cellular uptake was evaluated by fluorescence microscopy and flow cytometry. For the fluorescence microscopy experiments, cells were fixed with methanol containing 1 µg/mL DAPI to stain the cell nuclei, followed by three washing steps with PBS. Flow cytometry experiments were performed in 0.5% trypan blue (in PBS) to remove any extracellular fluorescence due to non-internalized nanoparticles. Even though FITC fluorescence is pH-dependent, given the short times employed in this experiment and the fact that nanoparticle intracellular distribution is expected to be similar among the different experimental groups, mean fluorescence intensity was employed to compare nanoparticle uptake between groups.

For cytotoxicity experiments, RGD-targeted and non-targeted nanoparticles were incubated with the cells (50 µg/mL suspension, 1 mL per well). Two hours after the addition of nanoparticles, an 808 nm NIR laser was applied in some of the wells (1 W/cm$^2$, 10 min irradiation) and cells were further incubated overnight. Then, the cells were rinsed with PBS, and cell viability was evaluated by Alamar Blue assay (Promega, Spain) following the manufacturer's instructions.

## 2.10 *Ex ovo* experiments

The chorioallantoic membrane (CAM) of chicken embryos grown *ex ovo* was employed as a model to evaluate anti-vascular therapeutic potential of the nanosystems prepared. Fertilized eggs were incubated from embryonic day (ED)-1 to ED-4 in a hatcher with rotation at 38 °C at 60 % humidity. On ED-4, the embryos were de-shelled following an established method[36] with the help of a Dremel drilling tool and transferred into sterilized weighing boats. A sterile square petri dish was used to cover the embryos, which were then transferred into a static humidified incubator at 38 °C, 60 % humidity and 0.5 % $CO_2$. Since chick embryos sacrificed before potential hatching, and experiments were carried out according to the available ethical guidelines, no especial institutional approval was necessary (as confirmed by the Animal Housing Service from UCM).

**2.10.1 Effect of released drugs on the CAM vasculature.** On ED-7, nanoparticles loaded with DOXY or a combination of DOXY and FROS (1 mg per sample) were suspended in PBS (0.5 mL) and stirred at 37 °C for 30 min. The particles were then centrifuged and the supernatant was employed to soak the cellulose disks that were placed on top of the CAM from chick embryos. PBS was also employed as a control. Three disks (each with one different experimental condition) were placed on each embryo. The vasculature around the disks was evaluated under a stereomicroscope and photographed using a 10x objective for the following 48 h. The blood vessels growing into the disk region from the surrounding area were counted at each time point. This number was normalized in regards to the initial number of vessels observed in the same region at the initial time.

**2.10.2 *Ex ovo* xenograft tumor model.** On ED-8, tumor xenografts were onplanted on the CAM chick embryos. In an Eppendorf tube, 0.4 mL of Collagen I (3 mg/mL) and 0.1 mL of 10x DMEM in cold (tubes introduced in a beaker with ice). Then, 2M NaOH was employed to neutralize the mixture (color change from the indicator in DMEM). Then, HT-1080 fibrosarcoma cells ($4x10^5$ cells) or a mixture of HT-1080 cells and dual drug-loaded nanoparticles (50 µg) in 50 µL total volume in 1x DMEM were added to the collagen mixture. Then, a slight damage was performed on the CAM by touching the surface of the CAM with tissue (close to a Y-shaped blood vessel) and 30 µL of the corresponding mixture was deposited in the area. Two onplants were placed on each CAM. Chick embryos were placed in the incubator for other 2 days, and the vascularization of the xenografts was evaluated under a stereomicroscope and photographed using a 10x objective on ED-10 (N=6). On ED-11, an 808 nm NIR laser was also applied on the vascularized onplants (2 W/cm$^2$, 5 min), and the effect was also evaluated under the stereomicroscope immediately after irradiation.

## 2.11 Statistical analysis

Two-tailed unpaired Students *t* test was employed to test for statistical differences among experimental groups. Differences were considered statistically significant for values of p<0.05.

## 3. Results and Discussion

Firstly, the integrity of all synthetized materials was checked by a complete characterization. The correct synthesis of gold nanorods (GNRs) was confirmed by Transmission Electron Microscopy (TEM) and UV-Vis-NIR spectrophotometry. Figure S3 shows a TEM micrograph of the prepared GNRs (showing their rod-like shape). As it was described previously, the reason for including GNRs within our mesoporous silica nanosystem is to heat the sample through NIR-light absorption of these plasmonic nanoparticles, inducing PTT. For biomedical applications, there is a spectral range in the near infrared between 650 and 900 nm (known as the first biological window), where the PTT is highly effective due to the enhanced penetration of this type of light through tissues. Thus, the laser used in our experiments is of 808nm, matching well the surface plasmon band of the prepared GNRs (maximum at a wavelength of 800nm, see UV-Vis-NIR in Figure S3).

The prepared core-shell nanoparticles (with GNR core and mesoporous silica shell, Au@MSNs) were characterized by different techniques (**Figure 2** and Figures S3-S6). Scanning electron microscopy (SEM) micrographs show round-shaped nanoparticles with a homogeneous size (Figure 2A). TEM micrographs show the presence of the gold nanorods surrounded by a porous silica shell (Figure 2B-C, Figure S3). Figure 2(D-F) also shows Dynamic Light Scattering (DLS) data with a hydrodynamic diameter around 90 nm and X-Ray Diffraction (XRD) patterns which confirm the presence of crystalline gold as well as an ordered mesoporous structure (at Small Angle measurements). Nanoparticle porosity was evaluated by Nitrogen adsorption, showing a surface area of around 300 $m^2/g$ and a pore diameter of 2.7 nm, in the mesopore range[37] (Figure S4). This initial Au@MSN

material was functionalized with amino groups on its external surface (Au@MSN-NH$_2$) to attach the photosensitizer ICG (Au@MSN-ICG) and eventually, a PEG chain without (Au@MSN-PEG) or with the targeting agent (Au@MSN-PEG-RGD). ICG was chemically coupled to the nanoparticles instead of physically adsorbed in the pores since ICG will retain its ROS-generating capacity when linked to the nanoparticle surface, as opposed to the drugs that will be loaded within the mesopores, since those have to be released from the material in order to exert their function. The presence of ICG in Au@MSN-ICG can be appreciated by the fluorescence of the material in the NIR region observed under fluorescence microscopy (Figure S4). The ICG grafting efficiency was 69.5 ± 4.9 % (as estimated by fluorimetry). The Fourier Transformed Infrared (FTIR) spectra of the materials confirm the correct functionalization with amino groups (bands *ca*. 1700 cm$^{-1}$ in Au@MSN-NH$_2$) and the proper grafting of PEG chains (bands 1400-1700 cm$^{-1}$ and 2800-3000 cm$^{-1}$ in Au@MSN-PEG-RGD) (Figure S5). A slight increase in nanoparticle size was observed after all these surface modifications (Au@MSN-PEG-RGD hydrodynamic diameter = 122.9 ± 17.7 nm), while their morphology and porous structure, as determined by TEM, were not affected (Figure S6). Furthermore, the presence of PEG in Au@MSN-PEG-RGD greatly enhanced suspension stability with 10% fetal bovine serum when compared to non-PEGylated Au@MSN (Figure S6D). Finally, Z Potential measurements show a negative charge in the as-prepared Au@MSNs, which changes to positive values for Au@MSN-NH$_2$, and slightly less positive values for Au@MSN-PEG, as was expected (Figure S5). All of these results from the physico-chemical characterization of the materials confirm the successful preparation of the desired nanostructures.

Once the materials were totally characterized, the properties of these new nanomaterials were evaluated one by one, starting with their selective targeting capabilities towards endothelial cells. Tumor vascular endothelial cells have been reported to over-express αβ-integrins, which interact with the RGD peptidic sequence that we have included in our nanodevice[26], while its expression in normal mature vasculature is minimal [38–40]. To evaluate the targeting capacities of our prepared nanosystems, fluorescein-labeled PEGylated nanoparticles were incubated *in vitro* with human umbilical vein endothelial cells (HUVEC, which also overexpress αβ-integrins that can interact with the RGD motif[28]). Fluorescence microscopy images as well as flow cytometry data show a dose-dependent nanoparticle uptake for both targeted and non-targeted nanoparticles (Figure S7, **Figure 3**). Nanoparticle internalization was much more efficient for targeted nanoparticles compared to non-targeted ones, at both concentrations tested (25 and 50 µg/mL), with a 3.7-fold increase in the mean fluorescence intensity at 25 µg/mL for Au@MSN-PEG-RGD compared to Au@MSN-PEG (fluorescence intensity histograms obtained by flow cytometry can be seen in Figure S7B). These data confirm that our prepared nanosystems will be endocytosed more efficiently in αβ-integrin-overexpressing cells, such as tumor vascular endothelial cells.

The next step was to evaluate the light-triggered responses that the nanosystems have been designed for, that is, their potential for PTT and PDT. The photothermal and photodynamic capacities of the synthesized nanosystems were first evaluated *in vial* (**Figure 4**). Suspensions of Au@MSNs and Au@MSNs-ICG (50 µg/mL) in 10 mM pH=7.4 phosphate-buffered saline (PBS) were exposed to NIR laser irradiation and both temperature and ROS generation were evaluated after stimulation. Both Au@MSNs and Au@MSNs-ICG were capable of inducing a temperature increase of similar magnitude, showing the good photothermal capacities of the GNRs in the system in the NIR range (Figure 4A). ROS generation upon the same NIR-light irradiation was corroborated only in the ICG-containing material (Figure 4B). This result highlights the necessity of all the components of the nanosystem to enable both photo-triggered behaviors (PTT and PDT), since the amount of ROS that could be generated from GNRs in our system was too low to be detected and, therefore, probably also too low to produce any biological effect.

In order to study the behavior of these light responsive nanomaterials in *in vitro* cell culture assays, the same nanosystems employed in the previous experiment were incubated with HUVEC vascular cells to test their effects on cell viability upon laser irradiation for 5 or 10 min (Figure S8, **Figure 5**). Neither laser irradiation without nanoparticles nor the incubation with the nanosystems in the absence of irradiation produced any significant decrease in cell viability. With 5 min NIR laser irradiation, significant decrease in cell viability was only observed for the cell treated with Au@MSN-ICG, but not for Au@MSN. This could indicate that the exposure of the cells to the generated temperature only for 5 minutes is not enough to produce any cell death, but its combination with ROS generation did produce a modest effect (81.2% mean cell viability). However, when both nanoparticles and 10 min laser irradiation were combined, a drastic decrease in cell viability was observed, which was even more pronounced for Au@MSN-ICG compared to Au@MSN (Figure 5). These results that: i) both the stimulus and the nanoparticles are well tolerated by vascular cells, ii) the developed nanodevices are capable of killing vascular cells when irradiated under appropriate conditions. In the complete system, the effects of PTT and PDT combined offer an improved behavior compared to relying merely on PTT alone.

After showing the NIR light-induced PTT and PDT capacities of our nanosystems, the remaining two anti-vascular modalities (anti-angiogenesis and disruption of neo-formed vessels) have to be provided by loading two drugs: fosbretabulin or FOS (a VDA)[33] and doxycycline or DOXY (an AAD)[32,41]. The combination of VDAs and AADs holds great promise and could remarkably improve the potential of antivascular treatments. FOS is a small-molecule VDA that induces selective vascular dysfunction and which is currently under clinical evaluation in several trials[22,33]. The chosen AAD, DOXY, has also been reported to be an effective drug against Cancer Stem Cells (CSCs), which would provide even another therapeutic mechanism to our nanoplatform. In fact, the hypoxic conditions that are generated in tumors subjected to antivascular therapies is known to drive CSC propagation and lead to the appearance of resistance mechanisms, limiting the efficacy of the treatment[42]. Therefore, including DOXY in our formulation could also prevent this CSC propagation, potentially contributing to the therapy[43,44].

After co-loading of both drugs, their release kinetics were evaluated *in vial* by fluorimetry (both drugs are fluorescent in different regions of the spectrum). The results from the drug release experiments can be found in **Figure 6**.

As is typical for mesoporous materials, the release kinetics of both drugs could be fitted to a First-order kinetic model (Equation 1).

$$Y=A(1-e^{-kt}) \tag{1}$$

where $Y$ the percentage of drug released at time $t$, $A$ the maximum amount of drug released (in percentage), and $k$, the release rate constant. The kinetic constant $k$ was larger for FOS than for DOXY, indicating a faster release for the former. Most of the loaded FOS was released within 4 h, while a similar percentage of DOXY release was achieved after 8 h. Danhier et al[45] reported significant accumulation of RGD-targeted nanoparticles in tumor endothelium just 2 h after intravenous injection, so it would be expected that the obtained nanosystem would still be co-releasing the chosen drugs when in the area of interest. The amount of loaded drug was also much larger for FOS than for DOXY (230.3 ± 5.1 µg/mg *vs* 23.2 ± 1.0 µg/mg of nanoparticles). It is interesting to note that when the drugs were loaded independently in Au@MSN instead of co-loaded, while no large differences were observed in DOXY loading (17.9 ± 0.4 µg/mg *vs* 23.2 ± 1.0 µg/mg), a large difference was found in FOS loading ( 52.3 ± 1.3 µg/mg *vs* 230.3 ± 5.1 µg/mg). We believe that the amount of FOS co-loaded with DOXY was greatly increased due to some interaction between both molecules (perhaps π-π stacking) that eased the retention of FOS inside the mesopores. Given the large difference in FOS loading, we decided to discard the sample loaded only with FOS for the following biological experiments, comparing then the sample loaded with DOXY and the one co-loaded with both drugs. Regarding the obtained release kinetics, the faster release of FOS could destroy the neo-formed vasculature of the tumor, entrapping more efficiently the nanoparticles in the region[19] and enabling the accumulation of DOXY, which would exert its antiangiogenic and anti-CSC activities for a longer period of time thanks to this initial destruction of tumor vasculature.

The effect of the drug-loaded materials was then evaluated employing an *ex ovo* chicken embryo chorioallantoic membrane (CAM) model. The *ex ovo* CAM model with its fast-developing vasculature enables a straightforward evaluation of pro- or anti-angiogenic responses when exposed to different stimuli, and permits performing several experiments per embryo. In this work, chick embryos were sacrificed before potential "hatching", so from the regulatory point of view, the bureaucratic process was greatly reduced compared to experimentation with murine models. Furthermore, this chick embryo model enabled the thorough evaluation of the biological effect of our nanodevice in a completely functional vascular system while also complying with the 3Rs principles[46]: replacing mammal use with chick embryos without fully developed nociceptive systems and reducing the number of animals needed (by evaluating several samples per embryo). To carry out the experiments, loaded-nanoparticles were suspended in PBS and left to release the drugs at 37 °C for 30 min. The particles were then centrifuged and the supernatant was employed to soak cellulose disks that were placed on top of the CAM from 7-day chick embryos. PBS was also employed as a control. The vasculature in the surrounding area was photographed for 2 days after disk placement and vessel growth was estimated by counting the small vessels contained in the area studied at the different time points. The results (Figure 7 and Figure S9) show a slowed vessel growth in the samples treated with DOXY compared to control, showing the antiangiogenic effect of the drug released from the nanoparticles. Even more interestingly, when both drugs were being released simultaneously, not only did vessel growth slow down, but an initial decrease in the number of small vessels was also observed, due to the effect of the vascular disrupting agent FOS (**Figure 7**).

The natural immunodeficiency of the chick embryo also enables the development of tumor xenograft models, which can be employed to evaluate the effect of nanoparticles intended for biomedical application[47]. To obtain tumor xenografts for evaluating the effect of our complete system, HT-1080 fibrosarcoma cells were onplanted on the CAM of chick embryos. Tumors without and with nanoparticles were onplanted to evaluate the effect of the different components of the nanosystem. Tumors with nanoparticles showed a significantly lower amount of blood vessels irrigating the tumors, due to the effect of the drugs being released from the nanoparticles (**Figure 8A**). Furthermore, while NIR irradiation had no visible effect on tumors without nanoparticles, in tumors containing nanoparticles, an evident hemorrhage due to vessel destruction by combined PTT and PDT was observed (**Figure 8B**).

The results here presented confirm the possibility of employing nanoparticles to exert multimodal antivascular therapeutics in a xenograft model, combining dual drug-induced impairment of the tumor vascular network, followed by the NIR light-induced destruction of the remaining vasculature.

## 4. Conclusions

In this work, we have presented a proof-of-concept strategy based on the development of nanoparticles with multimodal antivascular capabilities. This targeted nanoparticle based on a core/shell gold/mesoporous silica structure could simultaneously perform several functions: i) RGD-tagged nanoparticles were selectively internalized by αβ-integrin-overexpressing cells; ii) the nanoparticles where not toxic, but induced remarkable HUVEC cell death upon Near-Infrared laser irradiation; iii) two different drugs for anti-vascular therapy were co-loaded and released from the mesoporous silica shell; iv) the released drugs significantly affected the development of surrounding vasculature in a chick embryo model and v) the prepared nanosystem significantly decreased the number of blood vessels irrigating tumor xenografts *ex ovo,* and it also induced the destruction of existing blood vessels in the xenograft when exposed to Near-Infrared light.

**Conflicts of interest**

There are no conflicts to declare.

**Acknowledgements**

Funding from the European Research Council through the Advanced Grant VERDI (ERC-2015 AdG no.694160) is gratefully acknowledged. Images from Servier Medical Art (and their Powerpoint image bank) have been employed to produce Figure 1 in this article.

**References**

[1]    F. Bray, J. Ferlay, I. Soerjomataram, R.L. Siegel, L.A. Torre, A. Jemal, Global cancer statistics 2018: GLOBOCAN estimates of incidence and mortality worldwide for 36 cancers in 185 countries, CA. Cancer J. Clin. 68 (2018) 394–424.

[2]    H. Maeda, M. Khatami, Analyses of repeated failures in cancer therapy for solid tumors: poor tumor-selective drug delivery, low therapeutic efficacy and unsustainable costs, Clin. Transl. Med. 7 (2018) 11.

[3]    V.V. Padma, An overview of targeted cancer therapy, BioMedicine. 5 (2015) 19.

[4]    W.C.W. Chan, Nanomedicine 2.0, Acc. Chem. Res. 50 (2017) 627–632.

[5]    R. van der Meel, T. Lammers, W.E. Hennink, Cancer nanomedicines: oversold or underappreciated?, Expert Opin. Drug Deliv. 14 (2017) 1–5.

[6]    C.M. Dawidczyk, C. Kim, J.H. Park, L.M. Russell, K.H. Lee, M.G. Pomper, P.C. Searson, State-of-the-art in design rules for drug delivery platforms: Lessons learned from FDA-approved nanomedicines, J. Control. Release. 187 (2014) 133–144.

[7]    A. Baeza, E. Guisasola, E. Ruiz-Hernández, M. Vallet-Regí, Magnetically Triggered Multidrug Release by Hybrid Mesoporous Silica Nanoparticles, Chem. Mater. 24 (2012) 517–524.

[8]    R.X. Zhang, H.L. Wong, H.Y. Xue, J.Y. Eoh, X.Y. Wu, Nanomedicine of synergistic drug combinations

for cancer therapy – Strategies and perspectives, J. Control. Release. 240 (2016) 489–503.

[9] X. Xu, W. Ho, X. Zhang, N. Bertrand, O. Farokhzad, Cancer nanomedicine: from targeted delivery to combination therapy, Trends Mol. Med. 21 (2015) 223–232.

[10] T.L. Lin, L.F. Newell, R.K. Stuart, L.C. Michaelis, S.E. Rubenstein, H.S. Pentikis, T. Callahan, D. Alvarez, L.D. Mayer, A.C. Louie, CPX-351 ((Cytarabine:Daunorubicin) Liposome Injection, (Vyxeos)) Does Not Prolong Qtcf Intervals, Requires No Dose Adjustment for Impaired Renal Function and Induces High Rates of Complete Remission in Acute Myeloid Leukemia, Blood. 126 (2015).

[11] V.J. Venditto, F.C. Szoka, Cancer nanomedicines: so many papers and so few drugs!, Adv. Drug Deliv. Rev. 65 (2013) 80–8.

[12] J. Shi, P.W. Kantoff, R. Wooster, O.C. Farokhzad, Cancer nanomedicine: progress, challenges and opportunities, Nat. Rev. Cancer. 17 (2017) 20–37.

[13] S. Wilhelm, A.J. Tavares, Q. Dai, S. Ohta, J. Audet, H.F. Dvorak, W.C.W. Chan, Analysis of nanoparticle delivery to tumours, Nat. Rev. Mater. 1 (2016) 16014.

[14] T.T. Goodman, P.L. Olive, S.H. Pun, Increased nanoparticle penetration in collagenase-treated multicellular spheroids., Int. J. Nanomedicine. 2 (2007) 265–274.

[15] R. Carlisle, J. Choi, M. Bazan-Peregrino, R. Laga, V. Subr, L. Kostka, K. Ulbrich, C.-C. Coussios, L.W. Seymour, Enhanced Tumor Uptake and Penetration of Virotherapy Using Polymer Stealthing and Focused Ultrasound, JNCI J. Natl. Cancer Inst. 105 (2013) 1701–1710.

[16] C.D. Arvanitis, M. Bazan-Peregrino, B. Rifai, L.W. Seymour, C.C. Coussios, Cavitation-Enhanced Extravasation for Drug Delivery, Ultrasound Med. Biol. 37 (2011) 1838–1852.

[17] S. Barua, S. Mitragotri, Challenges associated with Penetration of Nanoparticles across Cell and Tissue Barriers: A Review of Current Status and Future Prospects., Nano Today. 9 (2014) 223–243.

[18] P. Bhattarai, S. Hameed, Z. Dai, Recent advances in anti-angiogenic nanomedicines for cancer therapy, Nanoscale. 10 (2018) 5393–5423.

[19] R. Jahanban-Esfahlan, K. Seidi, B. Banimohamad-Shotorbani, A. Jahanban-Esfahlan, B. Yousefi, Combination of nanotechnology with vascular targeting agents for effective cancer therapy, J. Cell. Physiol. 233 (2018) 2982–2992.

[20] M. Rajabi, S. Mousa, The Role of Angiogenesis in Cancer Treatment, Biomedicines. 5 (2017) 34.

[21] N. Ferrara, A.P. Adamis, Ten years of anti-vascular endothelial growth factor therapy, Nat. Rev. Drug Discov. 15 (2016) 385–403.

[22] D.W. Siemann, D.J. Chaplin, M.R. Horsman, Realizing the Potential of Vascular Targeted Therapy:

The Rationale for Combining Vascular Disrupting Agents and Anti-Angiogenic Agents to Treat Cancer, Cancer Invest. 35 (2017) 519–534.

[23] X.-Y. Wu, W. Ma, K. Gurung, C.-H. Guo, Mechanisms of tumor resistance to small-molecule vascular disrupting agents: Treatment and rationale of combination therapy, J. Formos. Med. Assoc. 112 (2013) 115–124.

[24] H. Xiong, Y. Wu, Z. Jiang, J. Zhou, M. Yang, J. Yao, pH-activatable polymeric nanodrugs enhanced tumor chemo/antiangiogenic combination therapy through improving targeting drug release, J. Colloid Interface Sci. 536 (2019) 135–148.

[25] J. Zhang, J. Li, Z. Shi, Y. Yang, X. Xie, S.M. Lee, Y. Wang, K.W. Leong, M. Chen, pH-sensitive polymeric nanoparticles for co-delivery of doxorubicin and curcumin to treat cancer via enhanced pro-apoptotic and anti-angiogenic activities, Acta Biomater. 58 (2017) 349–364.

[26] X. Li, M. Wu, L. Pan, J. Shi, Tumor vascular-targeted co-delivery of anti-angiogenesis and chemotherapeutic agents by mesoporous silica nanoparticle-based drug delivery system for synergetic therapy of tumor, Int. J. Nanomedicine. 11 (2015) 93.

[27] G. Villaverde, S. Gómez-Graña, E. Guisasola, I. García, C. Hanske, L.M. Liz-Marzán, A. Baeza, M. Vallet-Regí, Targeted Chemo-Photothermal Therapy: A Nanomedicine Approximation to Selective Melanoma Treatment, Part. Part. Syst. Charact. 35 (2018) 1800148.

[28] F. Yan, H. Wu, H. Liu, Z. Deng, H. Liu, W. Duan, X. Liu, H. Zheng, Molecular imaging-guided photothermal/photodynamic therapy against tumor by iRGD-modified indocyanine green nanoparticles, J. Control. Release. 224 (2016) 217–228.

[29] M. Vallet-Regí, A. Rámila, R.P. del Real, J. Pérez-Pariente, A New Property of MCM-41: Drug Delivery System, Chem. Mater. 13 (2001) 308–311.

[30] M. Vallet-Regí, E. Ruiz-Hernández, Bioceramics: from bone regeneration to cancer nanomedicine., Adv. Mater. 23 (2011) 5177–218.

[31] M. Vallet-Regí, M. Colilla, I. Izquierdo-Barba, M. Manzano, Mesoporous silica nanoparticles for drug delivery: Current insights, Molecules. 23 (2018) 1–19.

[32] M. Richardson, D. Wong, S. Lacroix, J. Stanisz, G. Singh, Inhibition by doxycycline of angiogenesis in the chicken chorioallantoic membrane (CAM), Cancer Chemother. Pharmacol. 56 (2005) 1–9.

[33] D.W. Siemann, D.J. Chaplin, P.A. Walicke, A review and update of the current status of the vasculature-disabling agent combretastatin-A4 phosphate (CA4P), Expert Opin Investig Drugs. 18 (2009) 189-197.

[34] M.N. Sanz-Ortiz, K. Sentosun, S. Bals, L.M. Liz-Marzán, Templated Growth of Surface Enhanced Raman Scattering-Active Branched Gold Nanoparticles within Radial Mesoporous Silica Shells, ACS Nano. 9 (2015) 10489–10497.

[35] J.S. Souris, C.H. Lee, S.H. Cheng, C.T. Chen, C.S. Yang, J. an A. Ho, C.Y. Mou, L.W. Lo, Surface charge-mediated rapid hepatobiliary excretion of mesoporous silica nanoparticles, Biomaterials. 31 (2010) 5564–5574.

[36] H.S. Leong, N.F. Steinmetz, A. Ablack, G. Destito, A. Zijlstra, H. Stuhlmann, M. Manchester, J.D. Lewis, Intravital imaging of embryonic and tumor neovasculature using viral nanoparticles, Nat. Protoc. 5 (2010) 1406–1417.

[37] J. Rouquerol, D. Avnir, C.W. Fairbridge, D.H. Everett, J.M. Haynes, N. Pernicone, J.D.F. Ramsay, K.S.W. Sing, K.K. Unger, Recommendations for the characterization of porous solids (Technical Report), Pure Appl. Chem. 66 (1994) 1739–1758.

[38] J.M. Albert, C. Cao, L. Geng, L. Leavitt, D.E. Hallahan, B. Lu, Integrin αvβ3 antagonist Cilengitide enhances efficacy of radiotherapy in endothelial cell and non–small-cell lung cancer models, Int. J. Radiat. Oncol. 65 (2006) 1536–1543.

[39] P.C. Brooks, A.M.P. Montgomery, M. Rosenfeld, R.A. Reisfeld, T. Hu, G. Klier, D.A. Cheresh, Integrin αvβ3 antagonists promote tumor regression by inducing apoptosis of angiogenic blood vessels, Cell. 79 (1994) 1157–1164.

[40] X. Chen, Multimodality imaging of tumor integrin alphavbeta3 expression., Mini Rev. Med. Chem. 6 (2006) 227–22734.

[41] L. He, A.G. Marneros, Doxycycline inhibits polarization of macrophages to the proangiogenic M2-type and subsequent neovascularization, J. Biol. Chem. 289 (2014) 8019–8028.

[42] E.M. De Francesco, M. Maggiolini, H.B. Tanowitz, F. Sotgia, M.P. Lisanti, Targeting hypoxic cancer stem cells (CSCs) with Doxycycline: Implications for optimizing anti-angiogenic therapy, Oncotarget. 8 (2017) 56126–56142.

[43] C. Scatena, M. Roncella, A. Di Paolo, P. Aretini, M. Menicagli, G. Fanelli, C. Marini, C.M. Mazzanti, M. Ghilli, F. Sotgia, M.P. Lisanti, A.G. Naccarato, Doxycycline, an Inhibitor of Mitochondrial Biogenesis, Effectively Reduces Cancer Stem Cells (CSCs) in Early Breast Cancer Patients: A Clinical Pilot Study., Front. Oncol. 8 (2018) 452.

[44] R. Lamb, B. Ozsvari, C.L. Lisanti, H.B. Tanowitz, A. Howell, U.E. Martinez-Outschoorn, F. Sotgia, M.P. Lisanti, Antibiotics that target mitochondria effectively eradicate cancer stem cells, across

multiple tumor types: Treating cancer like an infectious disease, Oncotarget. 6 (2015) 389–405.

[45] F. Danhier, B. Vroman, N. Lecouturier, N. Crokart, V. Pourcelle, H. Freichels, C. Jérôme, J. Marchand-Brynaert, O. Feron, V. Préat, Targeting of tumor endothelium by RGD-grafted PLGA-nanoparticles loaded with Paclitaxel, J. Control. Release. 140 (2009) 166–173.

[46] D.J. Wells, Animal welfare and the 3Rs in European biomedical research, Ann. N. Y. Acad. Sci. 1245 (2011) 14–16.

[47] B.T. Vu, S.A. Shahin, J. Croissant, Y. Fatieiev, K. Matsumoto, T. Le-Hoang Doan, T. Yik, S. Simargi, A. Conteras, L. Ratliff, C.M. Jimenez, L. Raehm, N. Khashab, J.-O. Durand, C. Glackin, F. Tamanoi, Chick chorioallantoic membrane assay as an in vivo model to study the effect of nanoparticle-based anticancer drugs in ovarian cancer, Sci. Rep. 8 (2018) 8524.

**Figure captions:**

**Figure 1.** Conceptual representation of the multimodal anti-vascular strategy developed in the present work.

**Figure 2.** Characterization of the prepared Au@MSNs: SEM micrograph (A), TEM micrographs (B,C), DLS data (D), XRD patterns at small (E) and conventional angles (F).

**Figure 3**. Nanoparticle uptake experiments of FITC-labeled Au@MSN-PEG and Au@MSN-PEG-RGD by HUVEC cells measured by fluorescence microscopy (top) and flow cytometry (bottom). Data are Means ±SD, N=3, **$p<0.01$, ***$p<0.001$.

**Figure 4**. *In vial* evaluation of temperature increase (A) and ROS generation (B) with Au@MSNs and Au@MSN-ICG upon NIR irradiation (808 nm, 1W/cm$^2$, 5 min). Temperature changes were determined by employing a fluorooptic probe, while ROS generation was evaluated by the fluorescent indicator H2DCFDA (As described in the Materials and Methods section). Data are Means ±SEM, N=3.

**Figure 5.** Cell viability data (measured by Alamar Blue assay) from HUVEC cells cultured with Au@MSN and Au@MSN-ICG without and with 10 min NIR laser irradiation (50 µg/mL nanoparticle suspension, 1 mL per well). Data are Means ±SD, N=3, ns=not significant, *$p<0.05$, **$p<0.01$, ***$p<0.001$.

**Figure 6.** *In vial* drug release experiments in PBS for FOS (left) and DOXY (right). Kinetic fitting to First-order model shown for each graph. Data are Means ±SD, N=3.

**Figure 7.** Effect on the vasculature of the CAM of chick embryos of cellulose disks soaked in the released media from DOXY-loaded and dual drug-loaded nanoparticles. Data are Means ±SD, N=3, *$p<0.05$.

**Figure 8.** Evaluation of the dual drug-loaded Au@MSN-PEG-RGD (NP group) and control experiments (without nanoparticle treatment) in a chick embryo tumor xenograft model. Effect of the nanoparticles on the vascularization of the onplanted xenografts (A) and effect of NIR irradiation on the existing vessels in and around the xenografts (B). Data are Means ±SD, N=6, *$p<0.05$. Scale bars represent 1 mm.

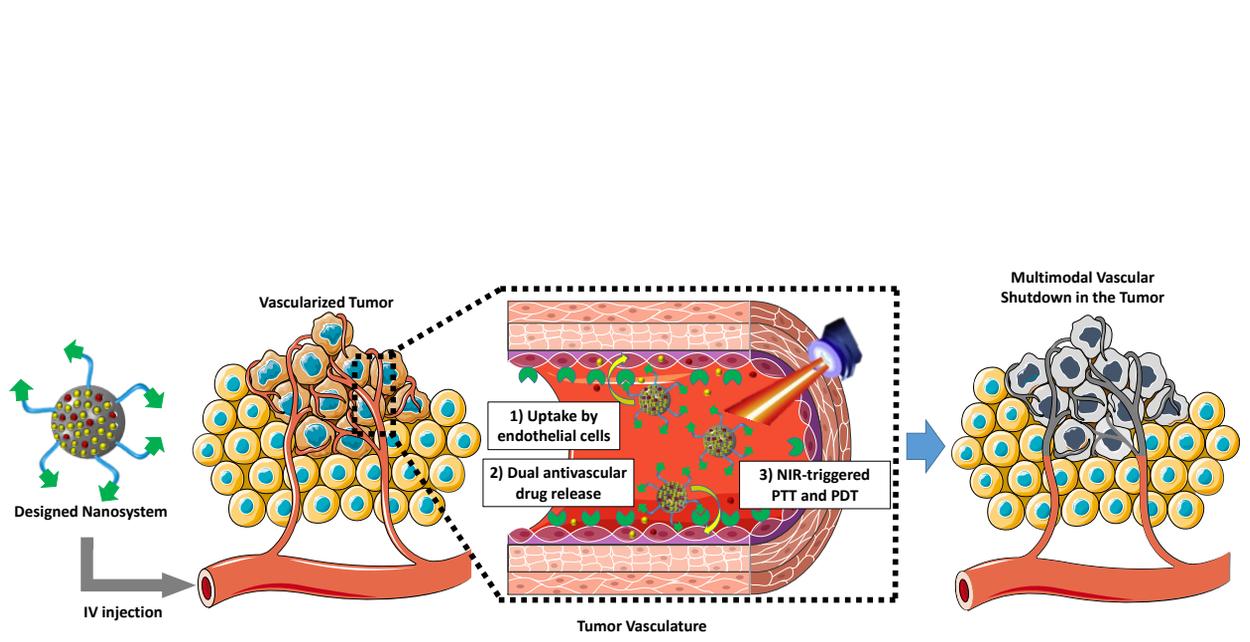

Figure 1.

**Figure 2**

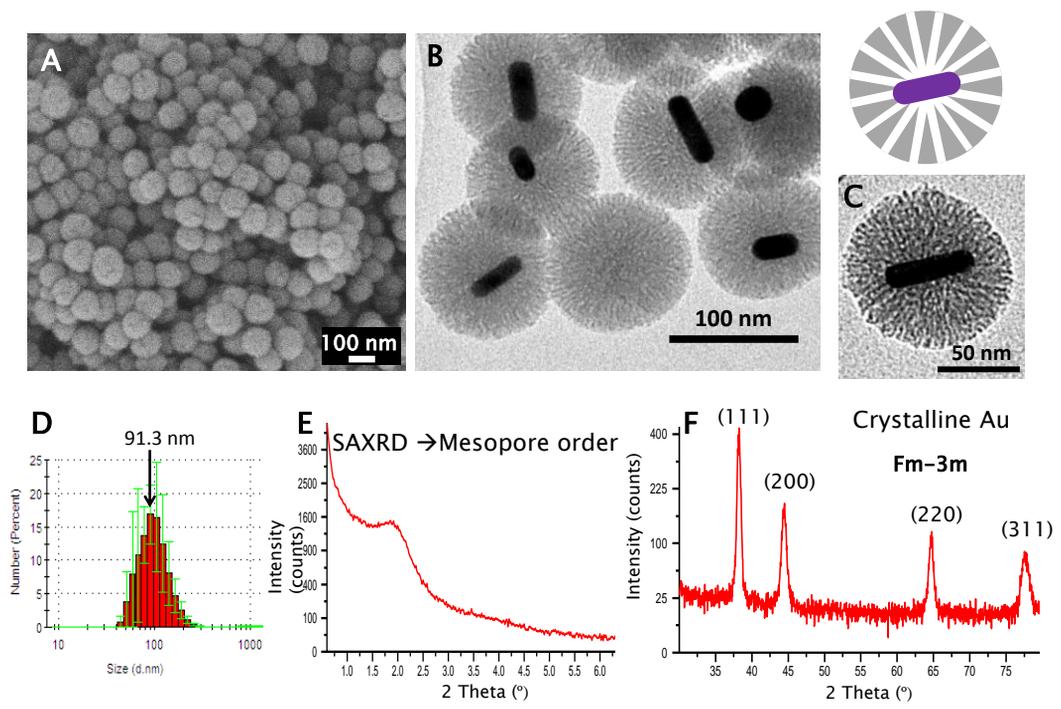

Figure 2.

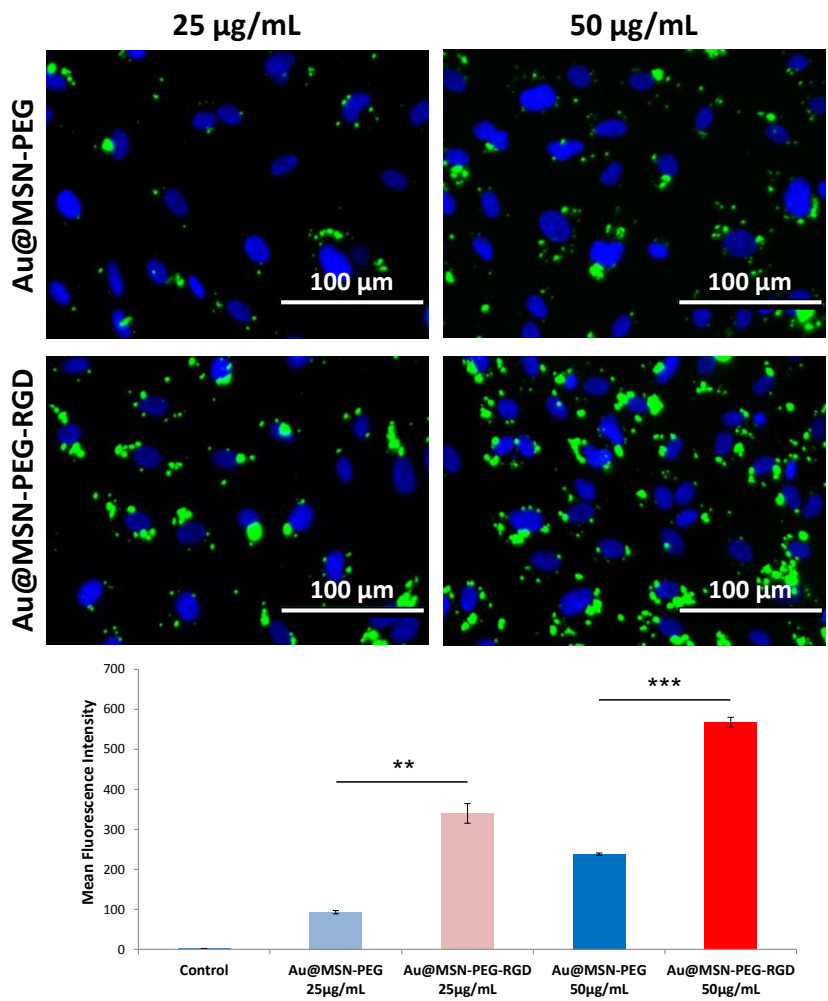

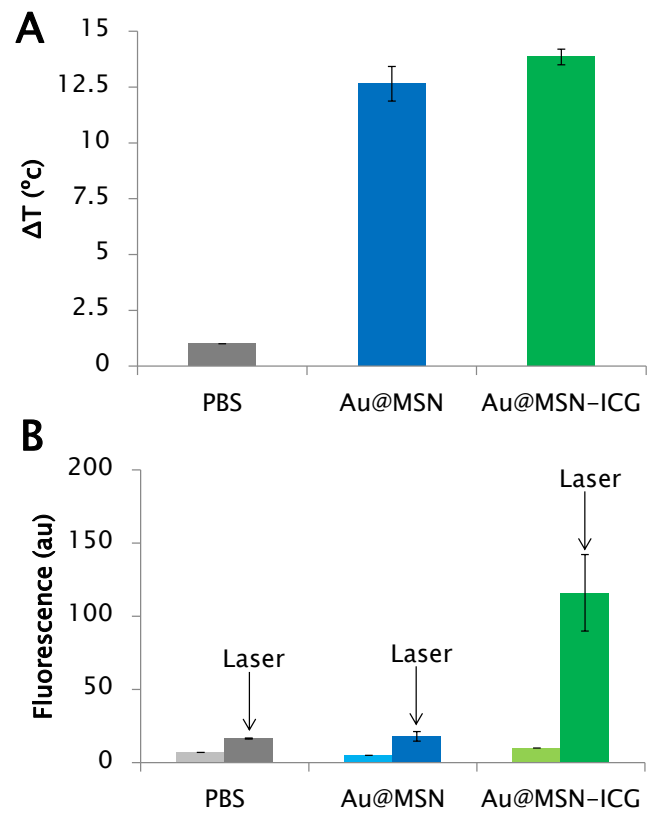

**Figure 4.**

**Figure 5**

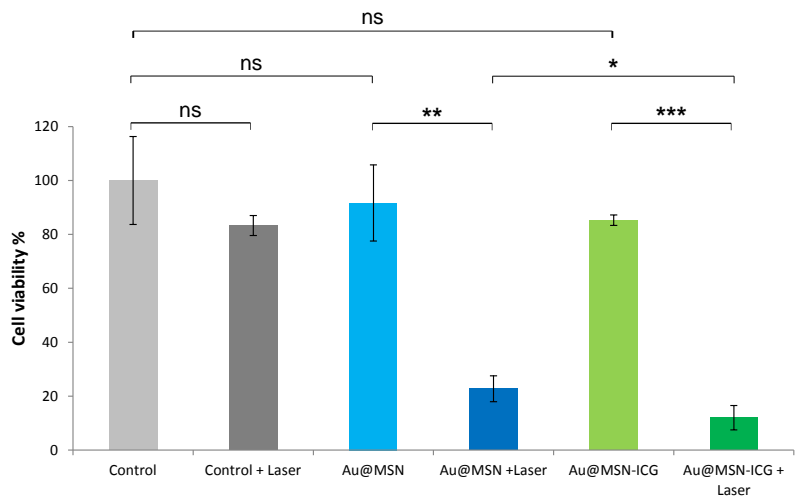

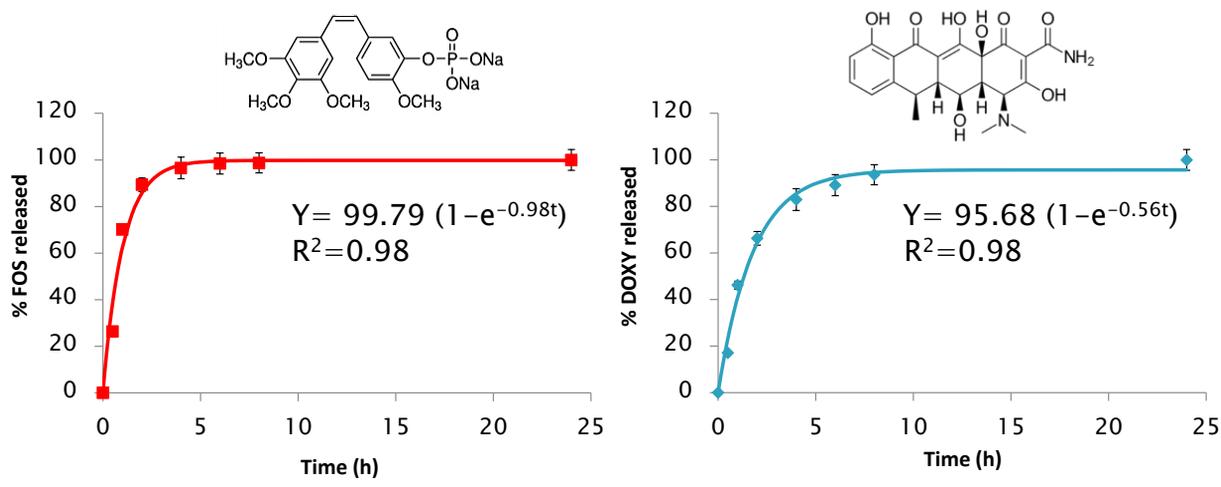

**Figure 7**

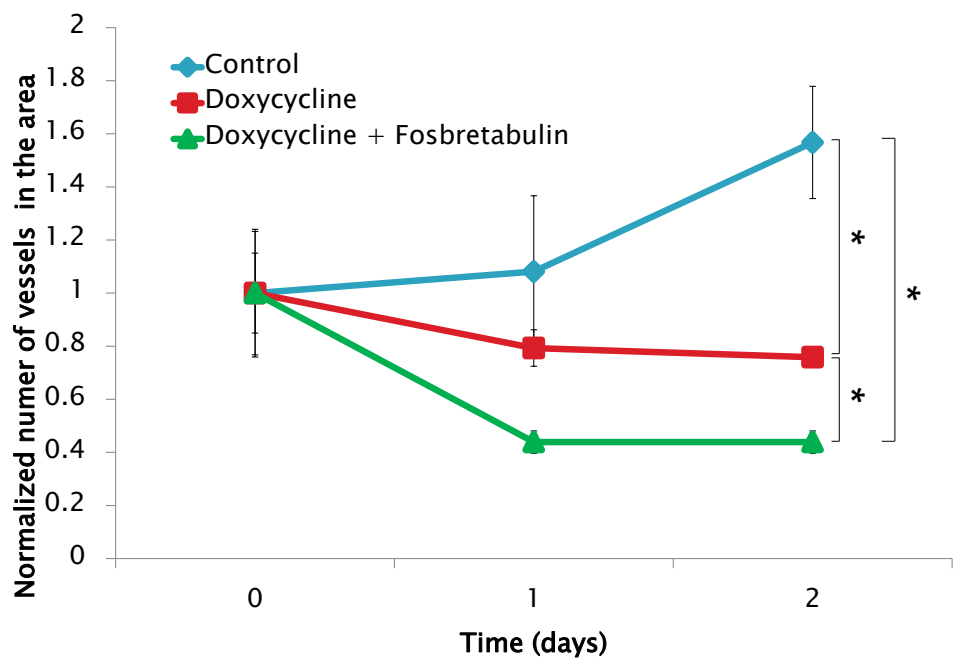

**Figure 7.**

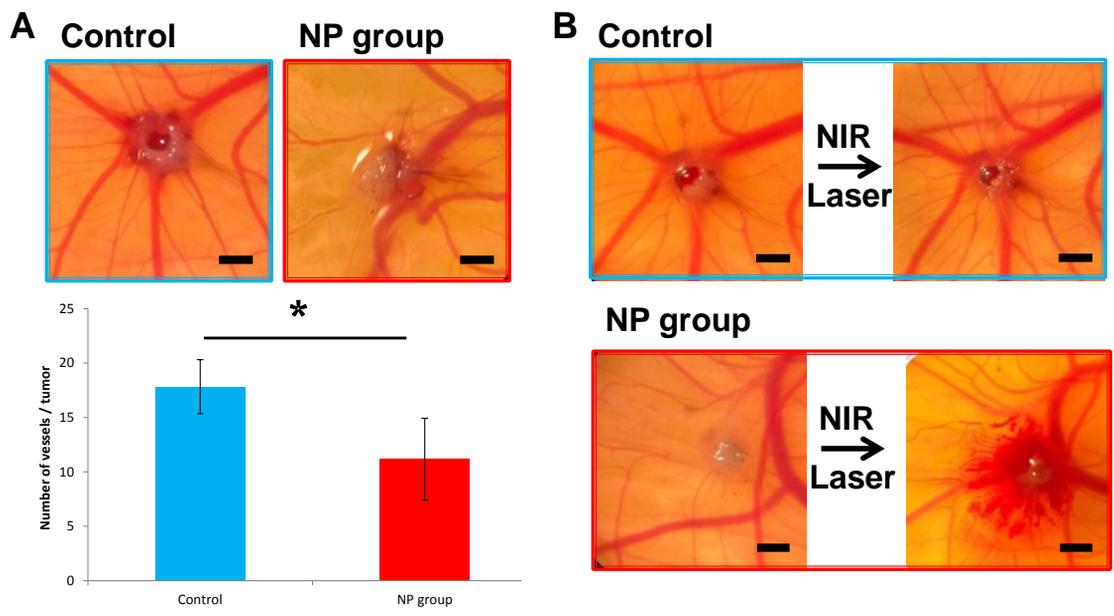